\documentclass[12pt]{iopart}

\usepackage{bm}
\usepackage{color}
\usepackage{epsfig}
\usepackage{citesort}
\usepackage{graphicx}
\eqnobysec

\newcommand{\lwig}{\mbox{\;\raisebox{.3ex}
    {$<$}$\!\!\!\!\!$\raisebox{-.9ex}{$\sim$}\;}}
\newcommand{\gwig}{\mbox{\;\raisebox{.3ex}
    {$>$}$\!\!\!\!\!$\raisebox{-.9ex}{$\sim$}}\;}
\newcommand{\lambdabar}{{\hbox{$\lambda$\kern-1.ex\raise+0.45ex\hbox{--}}}}

\begin{document}

\begin{flushright}
{\large \tt LAPTH-1263/08 \\
MPP-2008-88}
\end{flushright}

\title%
[Observing trans-Planckian ripples]{Observing trans-Planckian
ripples in the primordial power spectrum with future large scale
structure probes}

\author{Jan~Hamann}
\address{LAPTH (Laboratoire d'Annecy-le-Vieux de Physique
Th\'eorique, CNRS UMR5108 \&~Universit\'e de Savoie), BP 110, F-74941
 Annecy-le-Vieux Cedex, France}

\author{Steen~Hannestad, Martin~S.~Sloth}
\address{Department of Physics and Astronomy, University of Aarhus \\
Ny Munkegade, DK-8000 Aarhus C, Denmark}

\author{Yvonne~Y.~Y.~Wong}
\address{Max-Planck-Institut f\"ur Physik (Werner-Heisenberg-Institut) \\
F\"ohringer Ring 6, D-80805 M\"unchen, Germany}

\ead{\mailto{hamann@lapp.in2p3.fr}, \mailto{sth@phys.au.dk}, \mailto{sloth@phys.au.dk},
 \mailto{ywong@mppmu.mpg.de}}

\begin{abstract}
We revisit the issue of ripples in the primordial power spectra caused
by trans-Planckian physics, and the potential for their detection
by future cosmological probes. We find that for reasonably large
values of the first slow-roll parameter $\epsilon$ ($\gwig 0.001$), 
a positive detection of
trans-Planckian ripples can be made even if the amplitude is as low as
$10^{-4}$. Data from the Large Synoptic Survey Telescope (LSST) and
the proposed future 21 cm survey with the Fast Fourier Transform
Telescope (FFTT) will be particularly useful in this regard.  If the
scale of inflation is close to its present upper bound, a scale of new
physics as high as $\sim 0.2\ M_{\rm P}$ could lead to observable
signatures.
\end{abstract}

\maketitle

\section{Introduction}

Observations of the temperature and polarisation 
anisotropies in the cosmic microwave background (CMB)
radiation strongly support the idea that the initial conditions for
structure formation were set by an earlier period of cosmic 
inflation~\cite{Starobinsky:1980te,Guth:1980zm,Linde:1981mu,Albrecht:1982wi}.
During this period, the rapid expansion of space caused quantum
fluctuations of the inflaton field to be blown up to macroscopic
scales.  The initial state of these quantum fluctuations is typically
assumed to be given by the Bunch-Davies vacuum of de Sitter space at a
time when $k \gg aH$, i.e., the physical wavenumber $k/a$ is much
larger than the Hubble expansion rate.  Along with the usual assumptions of
single-field, slow-roll inflation, this leads to the almost
scale-invariant spectrum of Gaussian adiabatic scalar perturbations that
constitutes the ``primordial'' state of perturbations in today's
cosmological concordance model~\cite{Mukhanov:1990me}.

However, in an expanding space-time the notion of a vacuum is not
unique. This can easily be understood by observing that the expanding
space-time makes the Hamiltonian depend explicitly on time, so there
is no time-independent lowest energy state that can serve as a vacuum
at all times. We might then worry about how sensitive our predictions are to
the choice of initial state.  The choice of vacua other than the
Bunch-Davies one leads in principle to unacceptable ultraviolet
divergences in the observables~\cite{Brunetti:2005pr}, although this
deficiency can be cured by introducing a high-energy cutoff to the
theory.  In practice, such a cutoff will be given by the Planck scale,
or possibly even lower, by a \emph{scale of new
physics}~$\Lambda$ (if the cutoff is too close to the Planck scale, an
unorthodox vacuum choice may in fact lead to backreaction problems 
\cite{Tanaka:2000jw,Starobinsky:2001kn,Starobinsky:2002rp,KeskiVakkuri:2003vj,Porrati:2004gz,Greene:2004np}).
An ultraviolet cutoff in the inflaton fluctuation modes violates
energy conservation and has to be associated with a source term for
the fluctuations, in order to account for the modes redshifting across
the new physics hypersurface~\cite{KeskiVakkuri:2003vj}.

Unless inflation started only just before the wavelengths
corresponding to today's observable scales left the horizon, these
fluctuations can be mapped to wavenumbers larger than $\Lambda$ during
inflation and have emerged from above the new physics hyper-surface. 
It is thus very well conceivable that the fluctuations bear
an imprint of the unknown new physics~\cite{Martin:2000xs,Brandenberger:2000wr}. 
 In particular, the modes
might be created in a non-standard vacuum state at the new physics
hypersurface. Danielsson proposed a prescription of how to construct
this initial state by minimising the Hamiltonian on this hypersurface~\cite{Danielsson:2002kx}. 
However, it was found in
references~\cite{Bozza:2003pr,Giovannini:2003it} that the state
constructed in this fashion is somewhat ambiguous, since it is not
invariant under canonical transformations of the Hamiltonian. In fact,
it has been argued from a purely effective field theory point of view
that the effect on the primordial spectrum should be smaller than that
proposed by Danielsson~\cite{Kaloper:2002cs}.  Eventually, only a
better understanding of the thinning of ultraviolet degrees of freedom
in quantum gravity~\cite{Atick:1988si} can help us  solve the puzzle
of the correct choice of initial conditions and how the fluctuations
emerge during inflation, as in the proposed {\it trans-Planckian
damping} mechanism~\cite{Hassan:2002qk} (see also~\cite{Starobinsky:2002rp}).

Nonetheless, the generic signature of choosing a non-Bunch-Davies
vacuum state appears to be a superimposed oscillation of the primordial spectra
(``ripples''), whose
amplitude is suppressed by a power of $H/\Lambda$. For definiteness
we will consider Danielsson's model as a case study in the following.

The question of whether traces of these oscillations can possibly
be detected in present or future CMB anisotropy data has been the
subject of a number of recent studies
\cite{Bergstrom:2002yd,Elgaroy:2003gq,Martin:2003sg,Okamoto:2003wk,Martin:2004iv,Martin:2004yi,bib:ekp,Spergel:2006hy,Martin:2006rs,Groeneboom:2007rf}.
While few will disagree that the CMB anisotropies are at present
the most powerful probe of the primordial power spectrum,
the advent of high-redshift surveys of the
large scale structure (LSS) distribution
in the not-so-distant future may help to enlarge the detection window.
Planned galaxy redshift surveys such as the Large Synoptic Survey
Telescope (LSST) \cite{Ivezic:2008fe}, or surveys of the
distribution of neutral hydrogen using the 21~cm spin-flip line (e.g.,
the Murchison Widefield Array (MWA) \cite{bib:mwa}, the Square
Kilometre Array \cite{bib:ska}, or the Fast Fourier Transform
Telescope (FFTT) \cite{Tegmark:2008au}) have the advantage of
mapping out the density perturbations in three dimensions over
large volumes.
They are less affected by sampling variance
(or cosmic variance), which is ultimately the limiting factor for
CMB data over a large range of
scales.  In this work, we investigate how a combination of these
probes with CMB data can enhance our potential to discover the
footprints of new physics beyond the energy scale of inflation.

We shall begin by refreshing the reader's memory of the Danielsson
model in section~\ref{sec:model}. Some general considerations about
the phenomenology of oscillating spectra will be presented in
section~\ref{sec:precon}, before we turn to the questions of under which
conditions their traces can be detected in CMB data
(section~\ref{sec:detect?}), or, if no detection is possible, what
constraints can be placed on the parameters of the model
(section~\ref{sec:constraints}). In sections~\ref{sec:grs} and~\ref{sec:21cm}
we discuss how future galaxy redshift surveys and 21~cm
surveys can contribute to improving the chances for a detection of
ripples.  We conclude in section~\ref{sec:conclusions}.

\section{The model \label{sec:model}}

The spectrum of CMB perturbations predicted from inflation is usually
expressed in terms of the spectrum of co-moving curvature
perturbations, since this is a conserved quantity on super-Hubble
scales in single field slow-roll inflation. In the spatially flat
gauge, the curvature perturbation $\mathcal{R}(k)$ can be related
to the fluctuation of the inflaton field $\delta\phi(k)$ by the
relation $\delta\phi(k) = (\dot\phi/H)\mathcal{R}(k)$, where $H$ is
the Hubble scale and $\phi$ is the classical expectation value of the
inflaton field.

Let us remind the reader of the Danielsson initial condition and at
the same time generalise the derivation of the spectra to the case of
standard slow-roll inflation. Since there has been some confusion
about the correct generalisation to slow-roll in the literature, we
repeat some of the details.

In the spatially flat gauge we can write the metric in the form
\begin{equation}
ds^{2}=a^2(\tau)(d\tau^{2}-dx^{2}),
\end{equation}
and after transforming to the canonical variable $\mu = a\delta\phi$, the
perturbation of the inflaton field satisfies the Mukhanov equation,
which takes the form of the equation of motion of a harmonic
oscillator with a time-dependent mass \cite{Mukhanov:1990me},
\begin{equation}
\mu _{k}^{\prime \prime }+\left[ k^{2}-\frac{1}{\tau^2}\left(\nu^2-\frac{1}{4}\right)\right]
\mu _{k}=0,  \label{mukeq}
\end{equation}
where $\nu = 3/2+3\epsilon-\eta$, and $\epsilon$ and $\eta$ are the
usual slow-roll parameters which we assume to be constant.
 The conjugate momentum to $\mu _{k}$ is given by
\cite{Mukhanov:1990me}
\begin{equation}
\pi _{k}=\mu_k^{\prime }-\frac{z^{\prime }}{z}\mu _{k},
\end{equation}
where $z$ is defined as $z\equiv \sqrt{2\epsilon}a$.

Like for the ordinary harmonic oscillator, it is convenient to
quantise the system by promoting the fields to operators and expanding
them in terms of raising and lowering operators
\begin{eqnarray}
\hat \mu _{k}\left( \tau \right)  &=&\frac{1}{\sqrt{2k}}\left[\hat a_{k}\left( \tau
\right) +\hat a_{-k}^{\dagger }\left( \tau \right) \right],   \nonumber \\
\hat\pi _{k}\left( \tau \right)  &=&-i\sqrt{\frac{k}{2}}\left[\hat a_{k}\left( \tau
\right) -\hat a_{-k}^{\dagger }\left( \tau \right) \right] ,
\end{eqnarray}
where the vacuum is defined as the state annihilated by the lowering operator
\begin{equation}
\hat a_{k}\left(\tau\right)\left| 0 \right\rangle_{\tau} =0.
\end{equation}

Now, the vacuum at any later time is determined by a Bogoliubov transformation
\begin{eqnarray}
\hat a_{\bf k}\left( \tau \right)  &=&\alpha_{k}\left( \tau \right) \hat a_{\bf k}\left( \tau
_{0}\right) +\beta_{k}\left( \tau \right) \hat a_{\bf -k}^{\dagger }\left( \tau
_{0}\right),  \nonumber  \\
\hat a_{\bf k}^{\dagger }\left( \tau \right)  &=&\alpha_{k}^{*}\left( \tau \right)
\hat a_{\bf k}^{\dagger }\left( \tau _{0}\right) +\beta_{k}^{*}\left( \tau \right)
\hat a_{\bf -k}\left( \tau _{0}\right) ,
\end{eqnarray}
where $\alpha_k$ and $\beta_k$ are the Bogoliubov coefficients satisfying
 the normalisation condition
\begin{equation} \label{ab1}
|\alpha_k |^2-|\beta_k|^2=1,
\end{equation}
in order for the commutation relations to be preserved in time. The
solution to the dynamical equations is most easily found by defining a
set of mode functions
\begin{eqnarray}
f_{k}\left( \eta \right)  &=&\frac{1}{\sqrt{2k}}\left[ \alpha_{k}\left( \tau
\right) +\beta_{k}^{* }\left( \tau \right) \right],   \nonumber \\
g_{k}\left( \tau \right)  &=&-i\sqrt{\frac{k}{2}} \left[ \alpha_{k}\left( \eta
\right) -\beta_{k}^{*}\left( \tau \right) \right] ,
\end{eqnarray}
which satisfies the classical equation of motion, as can be seen by
employing the Heisenberg equation of motion. One can then invert these
relations to find
\begin{eqnarray}
\mu _{k}\left( \eta \right)  &=&f_{k}\left( \eta \right) a_{k}\left( \eta
_{0}\right) +f_{k}^{\ast }\left( \eta \right) a_{-k}^{\dagger }\left( \eta
_{0}\right),   \nonumber \\
\pi _{k}\left( \eta \right)  &=& g_{k}\left( \eta \right)
a_{k}\left( \eta _{0}\right) +g_{k}^{\ast }\left( \eta \right)
a_{-k}^{\dagger }\left( \eta _{0}\right).
\end{eqnarray}

By minimising the Hamiltonian written in terms of the variable $\phi$
and its conjugate momenta (note that minimising the Hamiltonian written
in terms of the Mukhanov variable gives a different result
\cite{Bozza:2003pr,Giovannini:2003it}), one can show that the
instantaneous vacuum $| 0 \rangle_{\tau_0}$ that minimises the energy
at $\tau=\tau_0$ is determined by
\begin{equation}
g _{k}\left( \tau _{0}\right) =-ik f _{k}\left( \tau _{0}\right) .
\end{equation}
This is equivalent to the zeroth order adiabatic vacuum, and the
condition is equivalent to requiring $\beta(\tau_0)=0$, which one can
use to normalise the general solution to the Mukhanov equation
\begin{eqnarray}
f_{k} &=&\frac{A_{k}}{\sqrt{2k}}e^{-i\zeta}\sqrt{-x}H_{\nu}^{(1)}(-x)
+\frac{B_{k}}{\sqrt{2k}}e^{i\zeta}\sqrt{-x}H_{\nu}^{(2)}(-x),
\nonumber \\
g_{k} &=&-A_{k}e^{-i\zeta}\sqrt{\frac{k}{2}}\sqrt{-x}H_{\nu-1}^{(1)}(-x)-B_{k}e^{i\zeta}\sqrt{\frac{k}{2}}
\sqrt{-x}H_{\nu-1}^{(2)}(-x),
\end{eqnarray}
where $A_{k}$ and $B_{k}$ are constants of integration to be fixed by
the physical initial conditions and $\zeta= -(\frac{1}{2}\nu+\frac{1}{4})\pi$. Solving the equations for the
Bogoliubov coefficients $\alpha_{k}$ and $\beta_{k}$, and expanding in
$1/x$, one obtains
\begin{eqnarray}
\alpha_{k} &=&\frac{1}{2}\left[ A_{k}e^{-ik\eta }\left( 2-i\frac{\nu^2-2\nu+1/2}{k\eta }\right)
+B_{k}e^{ik\eta }i\frac{\nu-1/2}{k\eta }\right],   \nonumber \\
\beta_{k}^{\ast } &=&\frac{1}{2}\left[ B_{k}e^{ik\eta }\left( 2+i\frac{\nu^2-2\nu+1/2}{k\eta }
\right) -A_{k}e^{-ik\eta }i\frac{\nu-1/2}{k\eta }\right] .
\end{eqnarray}

>From equation~(\ref{ab1}), we also have
\begin{equation}
\left| A_{k}\right| ^{2}-\left| B_{k}\right| ^{2}=1.
\end{equation}
As mentioned above, the Danielsson choice of vacuum requires that we
put $\beta_{k}\left( \tau _{0}\right) =0$ at some initial moment $\tau
_{0}$.  This implies that
\begin{equation}
B_{k}=ie^{-2ik\eta _{0}}\frac{\nu-1/2}{2k\eta _{0}+i(\nu^2-2\nu+1/2)}A_{k},
\end{equation}
from which we conclude that
\begin{equation}
\left| A_{k}\right| ^{2}=\frac{1}{1-\left| \chi _{k}\right| ^{2}},
\end{equation}
where
\begin{equation}
\chi _{k}=\frac{i(\nu-1/2)}{2k\eta _{0}+i(\nu^2-2\nu+1/2)}.
\end{equation}

One can then calculate the spectrum of curvature perturbations
$P_{\mathcal{R}}=(1/2\epsilon)P_{\phi}$, which becomes
\begin{eqnarray}
P_{\phi } &=&\frac{1}{a^{2}}P_{\mu }=\frac{k^{3}}{2\pi ^{2}
a^{2}}\left| f_{k}\right| ^{2}
\nonumber\\
&  \sim&  \frac{2^{2\nu-3}}{4\pi ^{2} \tau^{2}a^{2}} \left[\frac{\Gamma(\nu)}{\Gamma(3/2)}\right]^2(-k\tau)^{3-2\nu}\left( \left|
A_{k}\right| ^{2}+\left| B_{k}\right| ^{2}-A_{k}^{*
}B_{k}-A_{k}B_{k}^{* }\right)   \nonumber \\
&=&2^{2\nu-3}(1-\epsilon)^{2\nu-1} \left[\frac{\Gamma(\nu)}{\Gamma(3/2)}\right]^2\left( \frac{H}{2\pi }\right) ^{2}\nonumber\\
&  &\times \left( 1+\left| \alpha _{k}\right|
^{2}-\chi_{k}e^{-2ik\tau _{0}}-\chi_{k}^{*}e^{2ik\tau_{0}}\right)
\frac{1}{1-\left| \chi_{k}\right| ^{2}},
\end{eqnarray}
where we have taken the super-horizon limit $\tau\to 0$ and evaluated the
expression at horizon crossing $k=aH$, using 
$\tau
=-1/[aH(1-\epsilon)]$
in the prefactor.  Inserting $\chi$ gives,
up to a phase,
\begin{equation}
P_{\mathcal{R}}= P_{\mathcal{R}}^{BD}(k)\left[ 1-\frac{\nu-1/2}{x_0 }\sin(2x_0) \right],
\end{equation}
where $x_0=-k\eta_0$.

This expression agrees with the straightforward generalisation of the
corresponding calculation for power-law inflation in
references~\cite{Bozza:2003pr,Giovannini:2003it}, but disagrees
slightly with the result of \cite{Easther:2002xe}, who found that the
expression does not depend on $\eta$. This can be traced to an
inaccurate approximation.%
\footnote{The procedure outlined in the
footnote on page~7 of~\cite{Easther:2002xe} is slightly misleading.
Although the zeroth order adiabatic approximation is not exact, its
solution is only used as a boundary condition to normalise the exact
solution at the initial time giving the zeroth order adiabatic vacuum
solution, which is an exact solution. In the end we want to end up
with an exact solution, otherwise the effect would just be an effect
of an inaccurate approximation.}

Following the proposal of Danielsson, we should fix the initial
condition of a given mode when that mode crosses a fixed cutoff, i.e.,
\begin{equation}
|x_0| = \Lambda/H_0,
\end{equation}
where factors of order one can be absorbed in the arbitrary cutoff
$\Lambda$. Here $H_0$ is the Hubble parameter at $\tau=\tau_0$, when
the mode crosses the new physics hypersurface defined by the ultraviolet
cutoff $\Lambda$. Since $H_0$ is not the same for all  modes
because of the slowly decreasing  $H$, this will introduce $k$-dependent
oscillations in the spectrum. It can be shown that this leads to
$x_0\propto k^{\epsilon}$, since $H_0\propto k^{-\epsilon}$
\cite{Easther:2002xe}. One can then normalise $H$ to $H_0$ to obtain
\begin{equation}
|x_0| = \frac{\Lambda}{H_0}\left(\frac{k}{k_0}\right)^{\epsilon},
\end{equation}
which leads to
\begin{equation}
\label{eq:ripples}
P_{\mathcal{R}}(k) = P^{BD}_{\mathcal{R}}(k) \left( 1+(\nu-1/2) \xi \left( \frac{k}{k_0} \right)^{-\epsilon}
\sin \left[\frac{2}{\xi } \left(\frac{k}{k_0} \right)^\epsilon + \phi \right] \right),
\end{equation}
where $\xi \equiv H_0/\Lambda$. The phase $\phi$ needs to be added as
a free parameter since any dependence of the final result on the
(arbitrary) choice of pivot scale $k_0$ is non-physical and must
necessarily be spurious.  A similar expression exists for the tensor 
power spectrum, i.e., $P_h(k)=P^{BD}_h(k) ( \cdots)$.

\section{Preliminary considerations\label{sec:precon}}

In the limit $\epsilon \ll 1$, the argument of the sine function in
equation~(\ref{eq:ripples}) can be expanded in powers of $\epsilon$.
To the lowest order in $\epsilon$ and $\eta$,
\begin{equation}
\label{eq:osc}
P_{\mathcal{R},h}(k) \simeq P^{BD}_{\mathcal{R},h}(k) \left( 1+ \xi \left( \frac{k}{k_0} \right)^{-\epsilon}
\sin \left[\frac{2 \epsilon}{\xi} \ln \left(\frac{k}{k_0} \right) + \phi' \right] \right),
\end{equation}
i.e., the oscillations are in $\ln k$, with frequency $\omega \equiv 2
\epsilon/\xi$, or equivalently, oscillation length $L \equiv \pi
\xi/\epsilon$, and we have defined $\phi' \equiv
\phi+2(1+\epsilon)/\xi$.

\subsection{Constraints on the oscillation length \label{sec:osclength}}

The totality of CMB and LSS observations spans roughly four orders of
magnitude in $k$-space between $k_{\rm min} \sim 10^{-5} \ h \ {\rm
Mpc}^{-1}$ and $k_{\rm max} \sim 10^{-1} \ h \ {\rm Mpc}^{-1}$.  In
natural log space this range is approximately $\ln(k_{\rm max}/k_{\rm
min}) \simeq \ln(10^4) \simeq 9.2$.  For the individual experiment we
take the observable $k$ range to be that in which the measurements are
effectively noise-free and limited only by cosmic variance.

For a CMB survey that is cosmic variance-limited up to $\ell = 2000$
(the ``CVL survey''), we have $k_{\rm max} \simeq 0.15 \ h \ {\rm
Mpc}^{-1}$ within the concordance $\Lambda$CDM framework.  The Planck
satellite also has similar specifications in its $TT$
measurements~\cite{bib:planck}.  The WMAP $TT$ power spectrum is
cosmic variance-limited only up to $\ell \simeq
500$~\cite{Nolta:2008ih}, corresponding to $k_{\rm max} \simeq 5
\times 10^{-2} \ h \ {\rm Mpc}^{-1}$.  Therefore,
\begin{equation}
\label{eq:rcmb}
R_{\rm CMB} = \ln(k_{\rm max}/k_{\rm min})  \simeq \left\{\begin{array}{ll}
6.9, & {\rm WMAP}, \\
8.0, & {\rm Planck \ and \ CVL}, \\
\end{array}
\right.
\end{equation}
where in both cases we have used $k_{\rm min} \simeq 5 \times 10^{-5} \ h \ {\rm Mpc}^{-1}$.

If trans-Planckian corrections to the primordial power spectrum are to
be observed as an oscillatory signal, we must be able to fit in at
least three nodes inside the observable range $R_{\rm CMB}$.  Anything
less will not qualify as an oscillation, and the corrections will at
most show up as a spectral tilt or perhaps a running, or in the worst
case, as a normalisation.  Depending on the phase $\phi$ three nodes
is equivalent to $(1 \div 5/4) L$, where $L$ is the oscillation length
defined immediately after equation~(\ref{eq:osc}).  This then sets an
upper limit on the range of $\xi/\epsilon$,
\begin{equation}
\label{eq:upper}
\frac{\xi}{\epsilon} \lwig \frac{4}{5 \pi} R_{\rm CMB} \simeq \left\{\begin{array}{ll}
1.8, & {\rm WMAP}, \\
2.0, & {\rm Planck \ and \ CVL}, \\
\end{array}
\right.
\end{equation}
in which some vaguely oscillatory effect may be observable.

A lower bound on $\xi/\epsilon$ arises from the finite widths of the
CMB ``window functions''.  The CMB anisotropy spectra can be written
schematically as
\begin{equation}
C^{ij}_\ell  = \sum_{\alpha=s,t} \int d \ln k\  T^{i,\alpha}_\ell(k) T^{j,\alpha}_\ell (k) P_\alpha(k),
\end{equation}
where $i,j = T,E,B$, and $\alpha$ sums over scalar and tensor modes.
If the width of $T^{i,\alpha}_\ell(k) T^{j,\alpha}_\ell(k)$ is much
larger than the oscillation length $L$, then we expect the
oscillations to be wiped out, or, at least, suppressed by a factor of
a few.  For the $\Lambda$CDM model, the log width of
$T^{i,\alpha}_\ell(k) T^{j,\alpha}_\ell(k)$ at $\ell \sim 1000$ is
$\sim 0.07$, and increases with decreasing $\ell$.  Therefore, in
order to discern an oscillatory signature, we must have
\begin{equation}
\label{eq:lower}
\frac{\xi}{\epsilon} \gwig 0.022.
\end{equation}
We note that the window functions for the LSS power spectrum from
present galaxy redshift surveys are even broader.  For the power
spectrum of the Sloan Digital Sky Survey Luminous Red Galaxy (SDSS
LRG) sample published in reference~\cite{bib:lrg}, the window
functions have a log width of $\sim 0.33$, leading to $\xi/\epsilon
\gwig 0.1$. This means that present galaxy redshift surveys are not
very useful for resolving small ripples.

\subsection{Constraints on the oscillation amplitude \label{sec:amplitude}}

Suppose the true primordial power spectrum does have trans-Planckian ripples.  
One can then ask, how large does $\xi$ need to
be so that a smooth spectrum is a ``bad'' fit to the data.  This can
be estimated as follows.

The effective $\chi^2$ for a cosmic variance-limited CMB experiment in
one mode (e.g., $TT$) is given by
\begin{equation}
\chi_{\rm eff}^2 = f_{\rm sky} \sum_{\ell=2}^{\ell_{\rm max}} (2 \ell + 1) \left( \frac{C^{\rm obs}_\ell}{C_\ell}
+\ln \frac{C_\ell}{C_\ell^{\rm obs}} \right),
\end{equation}
where $C^{\rm obs}_\ell$ denotes the observed CMB anisotropy spectrum,
$C_\ell$ the theoretical predictions, and we have included a factor
$f_{\rm sky} \simeq 0.7$ to account for the inevitable sky cut.  The
difference in $\chi^2_{\rm eff}$ between the smooth and the
oscillatory models is then
\begin{equation}
\Delta \chi_{\rm eff}^2 = f_{\rm sky} \sum_{\ell=2}^{\ell_{\rm max}} ( 2 \ell+1) \left( \frac{C^{\rm obs}_\ell}{C^{\rm smooth}_\ell}
-\frac{C^{\rm obs}_\ell}{C^{\rm osc}_\ell}
+\ln \frac{C^{\rm smooth}_\ell}{C^{\rm osc}_\ell} \right),
\end{equation}
where $C_\ell^{\rm smooth}$ and $C_\ell^{\rm osc}$ denote the theoretical predictions
of the two respective models.

We express the oscillatory anisotropy spectrum $C_\ell^{\rm osc}$ in terms of the smooth one as
\begin{equation}
C_\ell^{\rm osc} = C_\ell^{\rm smooth} (1+\Delta_\ell),
\end{equation}
where $\Delta_\ell \ll 1$ contains the oscillatory features and in
general must be calculated by folding in the appropriate transfer
functions with the primordial power spectrum~(\ref{eq:ripples}).
Here, we estimate its effects by substituting $\Delta_{\rm rms}$ for
$\Delta_\ell$, where
\begin{equation}
\label{eq:delta}
\Delta_{\rm rms} = \frac{\xi}{\sqrt{2}}
\end{equation}
is the rms amplitude.
Then, keeping only terms to the lowest order in $\xi$,
we get
\begin{eqnarray}
\label{eq:chi2}
\Delta \chi_{\rm eff}^2 &\simeq & \frac{\xi^2}{4} f_{\rm sky} \sum_{\ell=2}^{\ell_{\rm max}} 2 \ell +1
\nonumber \\
&=&
\frac{\xi^2}{4} f_{\rm sky} (\ell_{\rm max}+3)(\ell_{\rm max}-1) \simeq \frac{\xi^2}{4} f_{\rm sky} \ell_{\rm max}^2,
\end{eqnarray}
where we have also used the expectation value $\langle C^{\rm obs}_\ell \rangle=C^{\rm osc}_\ell$.

As said, the expression~(\ref{eq:chi2}) assumes we are measuring in
one mode only (e.g., $TT$).  To account, approximately, for additional
modes (e.g., $TE$, $EE$), we need only to sum their contributions to
$\Delta \chi_{\rm eff}^2$.  We note that this procedure is strictly
not correct since the $TT$, $TE$ and $EE$ modes in principle
correlated, and a simple summation would tend to overestimate $\Delta
\chi_{\rm eff}^2$.  However, since we are interested only in order
magnitude estimates, a small overestimation is not a major concern.

Suppose our $TT$ measurements are cosmic variance-limited up to $\ell_{\rm max}=2000$ (e.g,
Planck).  Then
a ``$2 \sigma$'' detection of trans-Planckian ripples, i.e., $\Delta \chi^2_{\rm eff} \gwig 4$, would require
\begin{equation}
\label{eq:planck}
\xi \gwig 0.0024.
\end{equation}
If in addition $TE$ and $EE$  are cosmic variance-limited up to the same $\ell_{\rm max}$,
\begin{equation}
\label{eq:cvl}
\xi \gwig 0.0014.
\end{equation}
The corresponding limit for a WMAP-like measurement is
\begin{equation}
\label{eq:wmap}
\xi \gwig 0.0096,
\end{equation}
noting that the WMAP $TT$ measurement is effectively noise-free
only up to $\ell_{\rm max} \simeq 500$.

Finally, in deriving these amplitude limits we have assumed that $\xi$
is not degenerate with other cosmological parameters.  This is a good
approximation if $\xi$ satisfies the bounds~(\ref{eq:upper}) and
(\ref{eq:lower}) on the oscillation length.  In other words, if
trans-Planckian effects manifest themselves as oscillations in the
observable range, then no other parameter describing the primordial
power spectrum can mimic their effects on the cosmological
observables. Trans-Planckian ripples might conceivably be confused
with baryon acoustic oscillations (see reference~\cite{Covi:2006ci} for a
related scenario where this is the case).  We do not, however, find
this to be a problem since the ripples are oscillations in $\ln k$,
while the baryon acoustic oscillations are oscillations in
$k$. Results from our numerical parameter error forecasts corroborate
this claim.

\section{Can we detect ripples?\label{sec:detect?}}

\begin{figure}[t]
\includegraphics[width=15.5cm]{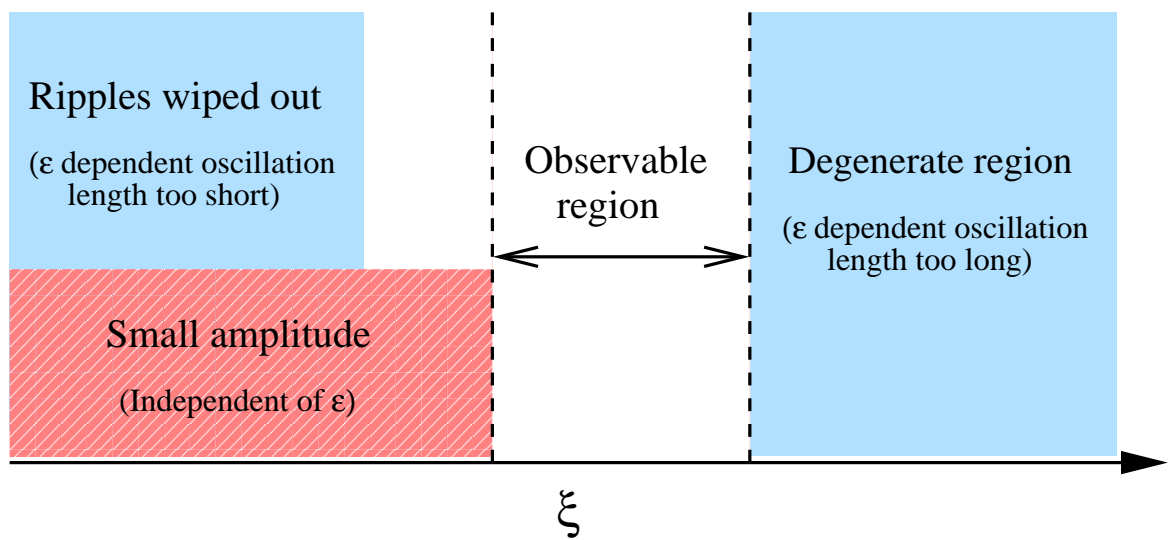}
\caption{Schematic showing the observability of trans-Planckian ripples.
The blue/solid box on the right represents the region in which the
oscillation length is so long that $\xi$ becomes degenerate with other
parameters describing the primordial power spectrum.
The blue/solid box on the left denotes the region in which the ripples
are wiped out by the window functions of the experiment because
of too short an oscillation length.  The red/hatched box on the left
encompasses the region in which the oscillation amplitude is too small
compared with the sampling error (i.e., cosmic variance).
Two vertical dashed lines demarcate the only window in which trans-Plankian
ripples may be detectable.\label{fig:schematic}}
\end{figure}

A simple picture emerges from the above considerations
(figure~\ref{fig:schematic}).  For a given $\epsilon$, oscillation
length requirements define a window in $\xi$ in which trans-Planckian
ripples are potentially observable.  Amplitude requirements define a
second, $\epsilon$-independent window.  Only when the two windows
overlap can one expect to see ripples.
Table~\ref{tab:table1} summarises the $\xi$ ranges in which
ripples can be observed for WMAP, Planck, and a CMB
survey that is cosmic variance-limited in $TT$, $TE$ and $EE$ up to
$\ell_{\rm max}=2000$ (CVL) for various fixed values of $\epsilon$.

Clearly, a crucial prerequisite for ripples detection is that
$\epsilon$ must be sizeable.  With $\epsilon$ fixed at a value as
large as $\gwig 0.01$, WMAP, Planck, and CVL all have the potential to
see trans-Planckian ripples, although the observable window for WMAP
is so small that taking into account the expected measurement error on
$\xi$,
\begin{equation}
\Delta \xi \sim \xi_{\rm min},
\end{equation}
where $\xi_{\rm min}$ denotes the minimum detectable $\xi$, would most
likely close it.  If we fix $\epsilon=0.001$, ripples
are unobservable for WMAP and Planck, but are in principle still in
range for CVL (barring the finite $\Delta \xi$).  No detection window
exists for $\epsilon=0.0001$ or smaller, since the associated
oscillation lengths are so large that $\xi$ always falls in the region
degenerate with the spectral tilt and/or the normalisation.

\begin{table}[t]
{\footnotesize
\caption{Ranges of $\xi$ in which trans-Planckian effects can be
observed as oscillations for WMAP, Planck, and CVL for fixed values of
$\epsilon$.  The parentheses denote those cases that are in reality
unobservable, either because of an undetectably small $\epsilon$, or
because the expected measurement error on $\xi$ exceeds the width of
the detection window.\label{tab:table1}}
\hspace{24mm}
\begin{tabular}{lccc}
\br
$\epsilon$ & WMAP  & Planck& CVL \\
\mr
0.02 & ($0.0096 \lwig \xi \lwig 0.036$) & $0.0024 \lwig \xi \lwig 0.041$ & $0.0014 \lwig \xi \lwig 0.041$ \\
0.01 & ($0.0096 \lwig \xi \lwig 0.018$) & $0.0024 \lwig \xi \lwig 0.020$  & $0.0014 \lwig \xi \lwig 0.020$ \\
0.001 & -- & -- & ($0.0014 \lwig \xi \lwig 0.0020$) \\
0.0001 & -- &--& --\\
\br
\end{tabular}
}
\end{table}

The question then is, how well do we expect to measure $\epsilon$ from
CMB data, since by the same arguments any measurement of $\epsilon$
that is consistent with $\epsilon=0$ means that no window exists for
the detection of trans-Planckian ripples.  A null detection of
$\epsilon$ by WMAP, $\epsilon \lwig 0.02 \ (95\% \ {\rm C.L.})$,
essentially rules out the observation of  ripples by
the current generation of CMB probes.  The projected $1 \sigma$
sensitivities to $\epsilon$ for Planck and CVL from our parameter
error forecasts (see \ref{sec:appendix} for details) are $\Delta
\epsilon \simeq 0.005$ and $\Delta \epsilon \simeq 0.0015$
respectively for an undetectable $\xi$.  Some variations exist if
there is a detection of $\xi$ (empirically, $\Delta \epsilon/\epsilon
\sim \Delta \xi/\xi$).  In table~\ref{tab:table1}, we enclose in
parentheses those scenarios that are most likely unobservable either
because of an undetectably small $\epsilon$, or because the expected
error on $\xi$ exceeds the width of the detection window.

\begin{figure}[t]
\includegraphics[width=15.5cm]{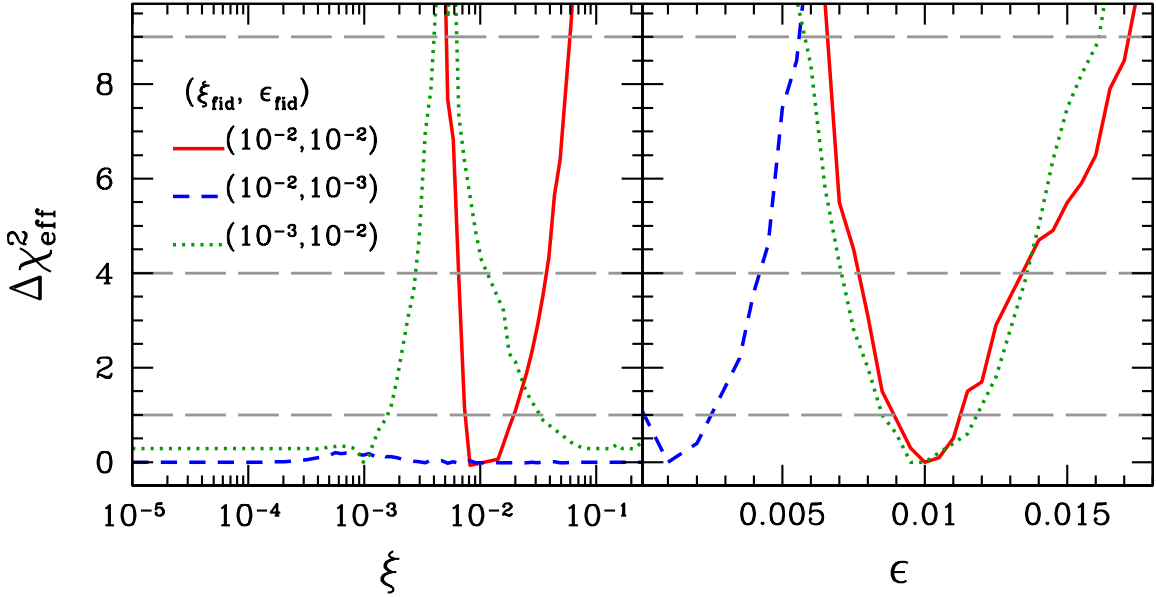}
\caption{$\Delta \chi^2_{\rm eff}$ curves  as functions of $\xi$ and
$\epsilon$ for three different fiducial models for the CVL experiment:
(i) $\xi_{\rm fid}=0.01$, $\epsilon_{\rm fid} =0.01$ (red/solid), (ii)
$\xi_{\rm fid}=0.01$, $\epsilon_{\rm fid} =0.001$ (blue/dashed), and
(iii) $\xi_{\rm fid}=0.001$, $\epsilon_{\rm fid} =0.01$
(green/dotted).  The three horizontal lines indicate, from top to
bottom, $\Delta \chi^2_{\rm eff}=9,4,1$ (i.e., $3 \sigma$, $2 \sigma$,
$1 \sigma$).
\label{fig:detection}}
\end{figure}

To confirm our analytic estimates we perform several rigorous
parameter error forecasts.  Figure~\ref{fig:detection} shows the
$\Delta \chi^2_{\rm eff}$ curves for three different fiducial models,
assuming the CVL experiment.  Technical details of the forecast and
definitions of the statistical quantities can be found in
\ref{sec:appendix}.

For fiducial model (i) with $\xi_{\rm fid} =0.01$ and $\epsilon_{\rm
fid}=0.01$, the prospects for ripples detection are excellent.  An
$\epsilon_{\rm fid}$ so large can be pinned down by CVL at better
than~$5 \sigma$.  This enables a detection of $\xi$ with a
correspondingly high statistical significance: between the best-fit
$\xi$ and at $\xi \to 0$ (i.e., a smooth spectrum), we find $\Delta
\chi^2_{\rm eff}\sim 40$.  However, if $\epsilon_{\rm fid}$ is lowered
to $0.001$ (model (ii)), no detection of $\xi$ is possible, since
$\epsilon=0.001 \pm 0.0015 \ (1 \sigma)$ is consistent with
$\epsilon=0$ so that $\xi_{\rm fid}=0.01$ lies in the degenerate
region.

Model (iii) with $\xi_{\rm fid}=0.001$ and $\epsilon_{\rm fid}=0.01$
is also futile, even though the projected $1 \sigma $ error on
$\epsilon$, $\Delta \epsilon \simeq 0.0015$, makes $\epsilon_{\rm
fid}$ well within experimental reach.  The reason here is that
$\xi_{\rm fid}=0.001$ leads to an unobservably small oscillation
amplitude.  Interestingly, although the $\xi_{\rm fid}$ of model~(iii)
is beyond CVL's reach, the very large and detectable $\epsilon_{\rm
fid}$ still makes it possible to exclude a range of $\xi$ values from
observations.  This is in contrast with model~(ii), for which an
undetectably small $\epsilon_{\rm fid}$ means that no conclusions can
be drawn about~$\xi$ (to be discussed in more detail in
section~\ref{sec:constraints}).

The results of figure~\ref{fig:detection} are in excellent agreement
with our analytic estimates presented in table~\ref{tab:table1}.
Using the same arguments we can also conclude that if $\xi_{\rm
fid}=0.01$ is to be detected by CVL, the underlying model must have
$\epsilon_{\rm fid} \gwig 0.005$.  This result is fully consistent
with a similar claim by Easther, Kinney and Peiris
(EKP)~\cite{bib:ekp}, who found that $\xi^{\rm EKP}=0.01$ is
detectable at $2 \sigma$ by a CMB experiment that is cosmic
variance-limited in $TT$, $TE$, $EE$ and $BB$ up to $\ell_{\rm
max}=1500$ only if $r \gwig 0.1$ (or $\epsilon \gwig 0.006$).%
\footnote{EKP used a power spectrum slightly different from our
equation~(\ref{eq:ripples}): $P(k) = P^{BD}(k) (\cdots)^{1/4}$.  This
means our $\xi/\epsilon$ constraints from oscillation length
considerations apply also to $\xi^{\rm EKP}/\epsilon$, while our lower
limits on $\xi$ from amplitude arguments need to be multiplied by a
factor of four for use with $\xi^{\rm EKP}$.}

\section{Can we constrain ripples?\label{sec:constraints}}

Suppose we detect $\epsilon=0.01$, and the true primordial power
spectrum has no trans-Planckian ripples.  Then using the same
amplitude arguments as those in section~\ref{sec:amplitude}, we can
put a $2 \sigma$ constraint on $\xi$ of
\begin{equation}
\label{eq:upperlimit}
\xi \lwig 0.0014
\end{equation}
for the CVL experiment.  If we take instead the EKP ripples and
experimental set-up, the same procedure gives a $2 \sigma$ limit of
$\xi^{\rm EKP} \lwig 0.0064$, which is in good agreement with EKP's
$\xi^{\rm EKP} \lwig 0.004$ for a scenario with $r=0.15$.

However, this upper limit is not the whole story.  This is because if
$\xi$ is very large ($\xi \gg \epsilon$), then for any given $\epsilon
\ll 1$, there exists an exact degeneracy between $\xi$ and~$\phi$.  To
see this we perform a small $x\equiv (2 \epsilon/\xi) \ln(k/k_0)$
expansion of the sine function in equation~(\ref{eq:osc}).  This gives
us a set of effective parameters
\begin{eqnarray}
\label{eq:effective}
A^{\rm eff}_{s,t} &=& A_s (1+\xi \sin \phi'), \nonumber \\
n^{\rm eff}_{s,t} &=& n_{s,t}  + \frac{\epsilon(2 \cos \phi' - \xi \sin \phi')}{1+\xi \sin \phi'}, \nonumber \\
\alpha^{\rm eff}_s &=& \alpha_s + \frac{2 \epsilon (n_s-1) (2 \cos \phi' - \xi \sin \phi')}{1+\xi \sin \phi'},
\nonumber \\
\alpha^{\rm eff}_t &=& \alpha_t + \frac{2 \epsilon n_t (2 \cos \phi' - \xi \sin \phi')}{1+\xi \sin \phi'}.
\end{eqnarray}
Applying the consistency relations $A_t^{\rm eff}/A_s^{\rm eff} = A_t/A_s = 16 \epsilon$ and
$n^{\rm eff}_t = n_t = - 2 \epsilon$, we see that
for whatever $\xi$ and $\epsilon$ we choose, setting $\phi'=\arctan(2/\xi)$
always returns the exact solutions $A_{s,t}=A^{\rm eff}_{s,t}$, $n_{s,t}=n^{\rm eff}_{s,t}$,
and $\alpha_{s,t}=\alpha^{\rm eff}_{s,t}$.

More (approximate) solutions exist if we ignore the consistency
constraints on $n_t^{\rm eff}$ and $\alpha^{\rm eff}_t$ (a good
approximation since future data are unlikely to resolve the tensor
power spectrum), or if the running of the scalar spectral tilt
$\alpha_s$ is treated also as a free parameter.  Thus, unless we have
a reason to fix $(\phi,n_s,A_s)$ to some particular values, it is
always possible to find a fit in the large $\xi \gg \epsilon$ limit
that is almost or even exactly as good as that of a smooth spectrum.
For this reason a lower limit on large $\xi$ should exist in addition
to the upper limit~(\ref{eq:upperlimit}) on small $\xi$.

Unfortunately there is no simple way to estimate this lower limit
accurately, since before we enter the exactly degenerate region at
$\xi \gg \epsilon$, there exists a region at $\xi \sim \epsilon$ in
which $\xi$ already exhibits some (partial) degeneracy with other
parameters describing the primordial power spectrum.  The small $x$
expansion limits the exactly degenerate region to $x \lwig 1$, or,
\begin{equation}
\label{eq:limit1}
\frac{2 \epsilon}{\xi} \ln\left(\frac{k}{k_0}\right) \sim \frac{2 \epsilon}{\xi} R_{\rm CMB} \lwig 1,
\quad \Rightarrow \quad \xi \gwig 18 \epsilon
\end{equation}
for the CVL experiment.  A less stringent limit can be set by the same oscillation length arguments
that led to equation~(\ref{eq:upper}), i.e.,
\begin{equation}
\label{eq:limit2}
\xi \gwig 2.0 \epsilon,
\end{equation}
again for CVL.  The true limit likely lies somewhere in between (\ref{eq:limit1}) and (\ref{eq:limit2}).

For $\epsilon\gwig0.01$, these lower limits on large $\xi$ are purely
academic since the trans-Planckian spectrum used here assumes $\xi\ll
1$.  They are important, however, in the case of small $\epsilon$,
since the lower limit~(\ref{eq:limit1}) or (\ref{eq:limit2}) may run
into the upper limit~(\ref{eq:upperlimit}), thereby closing the window
in which some finite $\xi$ may be a ``bad fit'' to the data.  For
example, no window exists for the case of a fixed $\epsilon=0.0001$.
In the same vein, if $\epsilon=0$ is consistent with data, then we can
draw no conclusions about $\xi$.

\begin{figure}[t]
\includegraphics[width=15.5cm]{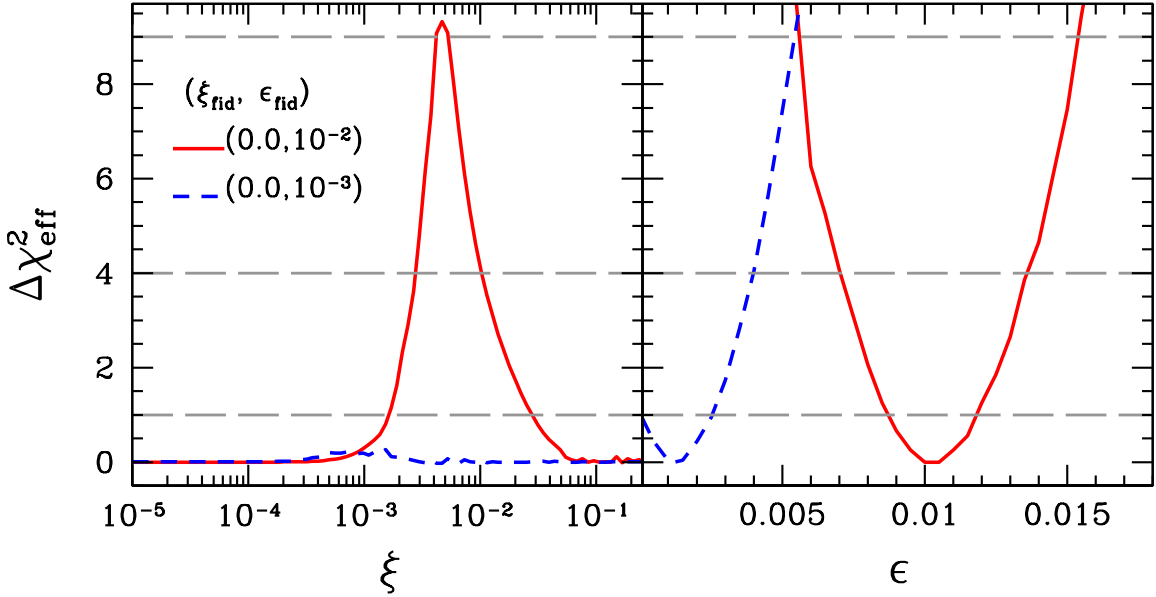}
\caption{Same as figure~\ref{fig:detection}, but for the models
(i) $\xi_{\rm fid}=0.0$, $\epsilon_{\rm fid} =0.01$ (red/solid), and (ii)
$\xi_{\rm fid}=0.0$, $\epsilon_{\rm fid} =0.001$ (blue/dashed).\label{fig:bounds}}
\end{figure}

Figure~\ref{fig:bounds} shows the $\Delta \chi^2_{\rm eff}$ curves for
two fiducial models, (i) $\xi_{\rm fid}=0.0, \epsilon_{\rm fid}=0.01$,
and (ii) $\xi_{\rm fid}=0.0, \epsilon_{\rm fid}=0.001$, assuming the
CVL experiment.  The bounds on $\xi$ for model (i) are in excellent
agreement with the predictions~(\ref{eq:upperlimit})
and~(\ref{eq:limit2}).  Model~(ii) has an undetectably small
$\epsilon_{\rm fid}$.  Correspondingly, no bounds can be set on $\xi$
in this case.

Lastly, we note that EKP found an upper limit on $\xi$ even in the
case of an undetectably small $\epsilon$, in apparent contradiction
with our results.  We believe this to be a consequence of the Bayesian
statistics used in their analysis.  In the small amplitude region at
small $\xi$, all values of the phase $\phi$ provide a good fit to the
data, since the oscillations are in any case unobservable.  In the
degenerate region at larger $\xi$ values, however, only certain
choices of $\phi$ give good fits.  This reduces the volume of the
posterior probability distribution in the $\phi$ direction at large
$\xi$ values, which are subsequently disfavoured in a Bayesian
analysis.

\section{Galaxy redshift surveys\label{sec:grs}}

Galaxy redshift surveys offer at present the most reliable probe of
the large scale structure distribution in the low-redshift universe
($z \lwig 1$).  Clustering statistics of galaxies from the catalogues
of the Two-Degree Field Galaxy Redshift Survey (2dFGRS) and the Sloan
Digital Sky Survey (SDSS) are now routinely used to constrain
cosmological parameters~\cite{bib:lrg,Cole:2005sx}.  Many
planned/proposed surveys will probe an even larger volume of the
universe and at higher redshifts in the future.  Here we consider the
potential of the Large Synoptic Survey Telescope (LSST; first light
2015)~\cite{Ivezic:2008fe}.

The LSST is a full-sky survey looking out to a maximum redshift of
$z_{\rm max} \sim 3$.  Within the $\Lambda$CDM framework this
corresponds to a survey volume of $V\sim 100 \ h^{-3} \ {\rm
Gpc}^{3}$.  Formally the noise-free region is defined as the range of
$k$ values for which the power spectrum signal $P_k$ dominates over
the Poisson shot-noise.  We take this region to be $k \lwig k_{\rm
max} \sim 0.2 \ h \ {\rm Mpc}^{-1}$.  We caution however that
nonlinear evolution such as mode-coupling may yet suppress the
trans-Planckian ripples or cause them to completely disappear beyond
some $k$ value smaller than $k_{\rm max}$.

The largest scale accessible to the LSST corresponds approximately to
$k_{\rm min} \simeq 2 \pi/V^{1/3} \simeq 1.4 \times 10^{-3} \ h \ {\rm
Mpc}^{-1}$.  The total $k$ range probed by the survey therefore spans
barely two orders of magnitude.  To maximise the spectral coverage we
combine the measurements from the LSST with those from a CMB
experiment.  Oscillation length arguments then impose an upper limit
on $\xi/\epsilon$ for which trans-Planckian ripples can be observed by
CMB+LSST,
\begin{equation}
\frac{\xi}{\epsilon} \lwig \frac{4}{5 \pi} \ln (k^{\rm LSST}_{\rm max}/k^{\rm CMB}_{\rm min}) \simeq 2.1,
\end{equation}
a number similar to the constraints~(\ref{eq:upper}) from the CVL
experiment alone.

The effective volume of the LSST is some 200 times larger than that of
the SDSS LRG sample.  Thus we can reasonably expect a corresponding
factor of $200^{1/3} \sim 6$ decrease in the widths of the $k$-space
window functions.  The window functions of the SDSS LRG power spectrum
has a log width of $\sim 0.33$, from which we derive a log width of
$\sim 0.055$ for the LSST.  This sets a lower limit on the oscillation
length of the trans-Planckian ripples, or, equivalently,
\begin{equation}
\frac{\xi}{\epsilon} \gwig 0.018,
\end{equation}
a limit comparable to the natural window functions of the CMB~(\ref{eq:lower}).

To estimate the lowest observable $\xi$ from amplitude arguments we
adopt the following expression for the effective $\chi^2$ of a
sampling variance-limited LSS survey~\cite{Tegmark:1997rp}:
\begin{equation}
\chi^2_{\rm eff} = \frac{V}{(2 \pi)^2} \int_{k_{\rm min}}^{k_{\rm max}} k^2 dk \left(\frac{P_k-P_k^{\rm obs}}{P_k} \right)^2.
\end{equation}
If the true primordial power spectrum has trans-Planckian ripples,
then the $\Delta \chi^2_{\rm eff}$ between a smooth and an oscillatory
spectrum is
\begin{equation}
\Delta \chi^2_{\rm eff} =  \frac{V}{(2 \pi)^2} \int_{k_{\rm min}}^{k_{\rm max}} k^2 dk \Delta_k^2,
\end{equation}
where $\Delta_k \ll 1$ encapsulates the oscillatory part of the power
spectrum.  As in section~\ref{sec:amplitude}, we substitute $\Delta_k$
with the rms amplitude $\Delta_{\rm rms} \equiv \xi/\sqrt{2}$, thus
leading to
\begin{equation}
\label{eq:lsstchi2}
\Delta \chi^2_{\rm eff} \simeq \frac{V}{(2 \pi)^2} \frac{\xi^2}{6} (k_{\rm max}^3-k_{\rm min}^3)
\simeq \frac{V}{(2 \pi)^2} \frac{\xi^2}{6} k_{\rm max}^3.
\end{equation}
Adding this $\Delta \chi^2_{\rm eff}$ to its CMB
counterpart~(\ref{eq:chi2}) yields the total $\Delta \chi^2_{\rm
total}$ for the combination CMB+LSST.

If we demand $\Delta \chi^2_{\rm total} \gwig 4$ for a $2 \sigma$ detection, then
\begin{equation}
\label{eq:lowerlimit}
\xi \gwig
\left\{\begin{array}{ll}
9.9 \times 10^{-4}, & {\rm Planck+LSST}, \\
8.5 \times 10^{-4}, &{\rm CVL+LSST}.\\
\end{array}
\right.
\end{equation}
Compared with the corresponding limit for Planck
alone~(\ref{eq:planck}), the projected constraint on $\xi$ for
Planck+LSST is more than a factor of two better.  Adding LSST to CVL
also yields some improvement over the limit from CVL
alone~(\ref{eq:cvl}).  Thus, with the LSST, we will be able to reach
down to the $\xi \sim 0.001$ level for the first time.
Table~\ref{tab:table2} summarises the ranges of $\xi$ in which
trans-Planckian ripples can be observed for Planck+LSST and CVL+LSST
for various fixed values of $\epsilon$.

\begin{table}[t]
{\footnotesize
\caption{Same as table~\ref{tab:table1}, but for
Planck+LSST, and CVL+LSST.\label{tab:table2}}
\hspace{24mm}
\begin{tabular}{lcc}
\br
$\epsilon$ & Planck+LSST  & CVL+LSST \\
\mr
0.02 & $9.9 \times 10^{-4}\lwig \xi \lwig 0.042$ & $8.5 \times 10^{-4} \lwig \xi \lwig 0.042$ \\
0.01 &  $9.9 \times 10^{-4} \lwig \xi \lwig 0.021$  & $8.5 \times 10^{-4} \lwig \xi \lwig 0.021$ \\
0.001 &  ($9.9 \times 10^{-4}\lwig \xi \lwig 0.0021$) & ($ 8.5 \times 10^{-4}\lwig \xi \lwig 0.0021$) \\
0.0001  &--& -- \\
\br
\end{tabular}
}
\end{table}

The remaining question is how sensitive are these combinations of CMB
and LSS probes to $\epsilon$.  In the case of a non-detection of
$\xi$, we find from our parameter error forecasts $1 \sigma$ errors of
approximately $\Delta \epsilon \sim 0.0015$ and $\Delta \epsilon \sim
0.005$ for Planck+LSST and CVL+LSST respectively.  These are
essentially the same sensitivities reachable by Planck and CVL alone.
In contrast, a detection of $\xi_{\rm fid}=0.01$ by Planck+LSST or
CVL+LSST does lead to some improvement in the sensitivity to
$\epsilon$, $\Delta \epsilon \sim 0.0007$, for $\epsilon_{\rm
fid}=0.01$.  In either case, $\epsilon=0.001$ remains out of reach.

\section{21~cm surveys\label{sec:21cm}}

Recently there is a growing interest to map the three-dimensional
distribution of neutral hydrogen via the observation of brightness
temperature fluctuations in the 21~cm spin-flip transition line.  A
number of radio arrays have been planned/proposed for this
observation, targeting specifically the redshift range $6 \lwig z
\lwig 20$~\cite{bib:mwa,bib:ska,Tegmark:2008au}.  At these redshifts
spatial fluctuations of the 21~cm signal are in general determined by
a combination of the underlying matter distribution and reionisation
physics.  However if the signal originates at a time when the neutral
hydrogen spin temperature is much larger than the CMB temperature
(i.e., $z \lwig 10$) and if at the same time the intergalactic medium
is completely neutral, the 21~cm spatial fluctuations are an exact
tracer of the perturbations in the underlying matter density field up
to a bias factor.  It is within this very optimistic scenario that we
consider the detectability of trans-Planckian ripples by 21~cm
surveys.  See reference~\cite{Furlanetto:2006jb} for a review on the
physics of the 21~cm transition.

Because they observe at high redshifts, 21~cm surveys tend to have
large effective volumes and can in principle probe larger wavenumbers
$k$ of the LSS power spectrum without impediments from nonlinear
evolution.  However, these surveys are limited at the low $k$ end of
the spectrum because of residual foregrounds.  Typically one expects
the foreground to swamp out the 21~cm signal at $k\lwig0.07 \ h \ {\rm
Mpc}^{-1}$.  Thus, it is necessary to combine 21~cm with CMB
observations to optimise $k_{\rm min}$.

To estimate $k_{\rm max}$ we use figure~3 of
reference~\cite{Pritchard:2008wy} which shows the effective volumes of
the Murchison Widefield Array (MWA)~\cite{bib:mwa}, the Square
Kilometre Array (SKA)~\cite{bib:ska}, and the Fast Fourier Transform
Telescope (FFTT)~\cite{Tegmark:2008au}.  We define $k_{\rm max}$ as
the point at which the effective volume drops below one half of the
maximum effective volume for each survey (effectively where the thermal noise of
the radio telescope becomes comparable to the 21~cm signal).  For MWA,
SKA and FFTT, we find $k_{\rm max} \sim 0.15, 0.6, 1 \ h \ {\rm
Mpc}^{-1}$ respectively.  Oscillation length arguments then lead
to the bounds
\begin{equation}
\frac{\xi}{\epsilon} \lwig \frac{4}{5 \pi} \ln (k^{\rm 21 cm}_{\rm max}/k^{\rm CMB}_{\rm min}) \simeq
\left\{\begin{array}{ll}
2.0, & {\rm CMB+MWA}, \\
2.4, & {\rm CMB+SKA}, \\
2.5, & {\rm CMB+FFTT}. \\
\end{array}
\right.
\end{equation}

To estimate the lowest observable $\xi$ by our 21~cm surveys we use
again the $\Delta \chi^2_{\rm eff}$ expression~(\ref{eq:lsstchi2}),
but replace $V$ with the effective volumes from figure~3 of
reference~\cite{Pritchard:2008wy} (i.e., $V \sim 1.5 \ h^{-3} \ {\rm
Gpc}^3$ for MWA and SKA, and $V \sim 150 \ h^{-3} \ {\rm Gpc}^3$ for
FFTT).  Again, we compute $\Delta \chi^2_{\rm total}$ by
adding $\Delta \chi_{\rm eff}^2$ to its CMB counterpart.
Demanding that $\Delta \chi^2_{\rm total} \gwig 4$ then gives
\begin{equation}
\xi \gwig
\left\{\begin{array}{ll}
0.0024, & {\rm Planck+MWA}, \\
0.0014, & {\rm Planck+SKA}, \\
7.9 \times 10^{-5}, & {\rm CVL+FFTT}. \\
\end{array}
\right.
\end{equation}
Table~\ref{tab:table3} contains a summary of the $\xi$ ranges detectable by
Planck+MWA, Planck+SKA, and CVL+FFTT.

\begin{table}[t]
{\footnotesize
\caption{Same as table~\ref{tab:table1}, but for Planck+MWA,
Planck+SKA, and CVL+FFTT.\label{tab:table3}}
\hspace{18mm}
\begin{tabular}{lccc}
\br
$\epsilon$ & Planck+MWA  & Planck+SKA  & CVL+FFTT \\
\mr
0.02 & $0.0024 \lwig \xi \lwig 0.040$ & $0.0014 \lwig \xi \lwig 0.048$ & $7.9 \times 10^{-5} \lwig \xi \lwig 0.050$ \\
0.01 & $0.0024 \lwig \xi \lwig 0.020$ & $0.0014 \lwig \xi \lwig 0.024$  & $7.9 \times 10^{-5} \lwig \xi \lwig 0.025$ \\
0.001 & --  & ($0.0014 \lwig \xi \lwig 0.0024$) & $ 7.9 \times 10^{-5}\lwig \xi \lwig 0.0025$ \\
0.0001 & -- &--&  ($ 7.9 \times 10^{-5}\lwig \xi \lwig 2.5 \times 10^{-4}$)   \\
\br
\end{tabular}
}
\end{table}

With a potential to reach down to $\xi \lwig 10^{-4}$, CVL+FFTT
improves the sensitivity of CVL alone by a factor of 20.  That CVL and
FFTT observe at different wavenumbers (but with some overlap) also
allows their combination to probe, for some given~$\epsilon$, larger
values of $\xi$ than either experiment alone.  Of particular note is
that the improved limits will create, for the first time, a sizeable
detection window for $\epsilon$ values as small as~$0.001$.

\section{Conclusions \label{sec:conclusions}}

We have investigated the sensitivity of various future cosmological
probes to ``ripples'' in the primordial power spectrum due to
``trans-Planckian'' physics.  The magnitude of this effect is
determined by $\xi=H_0/\Lambda$, i.e., the energy scale of inflation
relative to the scale of new physics. As also noted in some previous works on 
the topic (e.g., \cite{Elgaroy:2003gq}), 
the sensitivity to $\xi$ of a given probe is determined by
two experimental parameters.  One is the precision with which the power
spectrum can be measured over the relevant range, specified by the error bars
of the data and/or the width of the window functions.  The other
important parameter is the ``lever arm'', i.e., the span in $k$-space
over which the power spectrum can be measured, which is crucial for
disentangling  ripples in the power spectrum caused by
trans-Planckian physics from a spectral tilt or an amplitude shift.

Qualitatively, the precision of the data is the limiting factor in the
case of very small~$\xi$, while the lever arm determines, for a given $\epsilon$, the largest
$\xi$ that can be detected uniquely as ripples in the primordial power spectrum. 
The limit imposed by the lever arm also means that the detectability of
trans-Planckian ripples depends strongly on the magnitude of the first
slow-roll parameter $\epsilon$. If $\epsilon$ is too small, 
no detection window exists for $\xi$, 
as can be seen from
equation~(\ref{eq:osc}), as well as more quantitatively from tables~\ref{tab:table1}
 to~\ref{tab:table3}.  In the same vein, if $\epsilon=0$ is consistent with data, 
then no bounds can be set on $\xi$.

We have estimated the detection threshold of these ripples for data
from future CMB experiments (Planck and a hypothetical cosmic variance
limited experiment CVL), a galaxy redshift survey exemplified by
the LSST, and finally a large scale 21 cm survey. Due to their very
low noise and large volumes, both galaxy and 21 cm surveys
will eventually reach a higher precision than CMB observations alone. 
This will allow us to reach down to smaller values of $\xi$ than ever before.
However, large scale structure surveys will not be able to compete with 
 CMB observations in terms of wavelength coverage.  
 Therefore, in order to maximise the lever arm for the detection 
of trans-Planckian ripples, these surveys should be 
combined with CMB measurements.

In an optimistic scenario with $\epsilon=0.01$, we find that on the intermediate timescale, 
data from Planck+LSST will be open a
detection window spanning $10^{-3} \lwig
\xi \lwig 0.02$.
In the more distant future the lower limit of the window 
can possibly be pushed down to 
$10^{-4}$ using data from a cosmic variance limited CMB experiment combined 
with the FFTT 21 cm survey.  
This same combination of data will also create a window of detection 
for $\epsilon=0.001$ and possibly lower for the first time 
($10^{-4} \lwig \xi \lwig 0.0025$).

So what are the implications for the scale of new physics $\Lambda$,
which is presumably close to or at the Planck scale?  Given that
present data already constrains the scale of inflation to $H_0 \lwig 2
\times 10^{-5}\ M_{\rm P}$~\cite{Valkenburg:2008cz}, we conclude that
in the Danielsson model, there are no realistic prospects of ever
detecting the traces of new physics in the primordial spectra if
$\Lambda \gwig 0.2\ M_{\rm P}$.  For models in which the amplitude of
the ripples is quadratic instead of linear in $\xi$, as predicted by
purely effective field theory arguments in reference~\cite{Kaloper:2002cs},
the new physics would need to set in at energies smaller than $ \sim 2
\times 10^{-3}\ M_{\rm Pl}$.

Nonetheless, we have shown that the fantastic precision of future
CMB surveys like Planck, and large scale structure surveys such as LSST and FFTT 
will extend the detection window of trans-Planckian ripples by
more than two orders of magnitude compared with present data. 
After all, the prospect of probing physics
close to the Planck scale is a most tantalising one; a detection of
trans-Planckian signatures would truly be one of the most important
accomplishments of modern precision cosmology.

\section*{Acknowledgements}

We acknowledge use of computing resources from the Danish Center for
Scientific Computing (DCSC). The work of JH was supported by the ANR
(Agence Nationale de la Recherche).

\appendix

\section{Details of the parameter error forecast\label{sec:appendix}}

The cosmological model used in our forecast consists of nine free
parameters, the fiducial values of which are as follows: the baryon
density $\Omega_b h^2 = 0.02273$, dark matter density $\Omega_c h^2 =
0.1099$, Hubble parameter $h=0.72$, redshift to reionisation $z_{\rm
re}=12$, scalar spectral index $n_s=0.96$ and normalisation $\ln(
10^{10} A_s) = 3.18$.  The fiducial values of the slow-roll parameter
$\epsilon$ and the ripple parameter $\xi$ are given in the main text
of the paper, while the fiducial phase $\phi$ is 0.  In addition, we
use the following consistency relations on the tensor spectral index
$n_t$ and normalisation $A_t$: $n_t=-2 \epsilon$, and $A_t/A_s= 16
\epsilon$.

We generate mock CMB data using the method discussed in
reference~\cite{Perotto:2006rj}.  The likelihood function ${\cal L}$ is
defined as
\begin{equation}
\label{eq:likelihood}
\chi^2_{\rm eff} \equiv -2 \ln {\cal L} = \sum_{\ell=2}^{\ell_{\rm max}} (2 \ell +1) \ f_{\rm sky}
\left[{\rm Tr}(\widetilde{\bm C}_\ell^{-1} \hat{\bm C}_\ell) +
\ln \frac{|\widetilde{\bm C}_\ell|}{|\hat{\bm C}_\ell|} - n \right],
\end{equation}
where $\hat{\bm C}_\ell \doteq \hat{C}^{\mu \nu}_\ell$, $\mu,\nu=
T,E$, denotes the mock data covariance matrix, $\widetilde{\bm C}_\ell
\doteq C^{\mu \nu}_\ell + N^{\mu \nu}_\ell$ is the total covariance
matrix comprising theoretical predictions of the CMB anisotropy
spectrum $C^{\mu \nu}_\ell$ and the noise power spectrum $N^{\mu
\nu}_\ell$.  The quantity $n$ counts the number of observable modes.
For observations in temperature and $E$-type polarisation, $n=2$.

Parameter estimation is performed by $\chi^2$-minimisation, using a
simulated annealing routine coupled to the Boltzmann code
CAMB~\cite{Lewis:1999bs}.  We often plot $\Delta \chi^2_{\rm eff}(x)$
as functions of the parameters we wish to constrain (e.g., $x=\xi,
\epsilon$).  This is defined as
\begin{equation}\label{eq:deltachi2}
 \Delta \chi^2_{\rm eff}(x)
 \equiv - 2 \ln\left[
 \frac{{\cal L}^{(1)}(x)}
 {{\cal L}_{\rm max}}\right],
\end{equation}
where ${\cal L}_{\rm max}$ is the global
maximum of the likelihood function ${\cal L}$  defined in~(\ref{eq:likelihood}), and
\begin{equation}
{\cal L}^{(1)}(x) \propto
\max_{y_1,\ldots,y_N}  {\cal L}(x,y_1,\ldots,y_N)
\end{equation}
is a projection of  ${\cal L}(x,y_1,\ldots,y_N)$
onto the one-dimensional subspace $x$ by maximising along the
$y_1,\ldots,y_N$ directions.

We define our ``$1 \sigma$'', ``$2 \sigma$'' and ``$3 \sigma$''
intervals as the regions of $x$ satisfying respectively $\Delta
\chi^2_{\rm eff} \leq 1$, $4$, and $9$.  We
emphasise that these intervals have no formal probabilistic
interpretation. However, if ${\cal L}^{(1)}(x)$ is a Gaussian
distribution and we assume flat priors on the model parameters, then
these intervals are exactly identical to the 68\%, 95\%, and 99\%
credible intervals derived from a Bayesian analysis.  See
reference~\cite{Hamann:2007pi} for more details on Bayesian versus
non-Bayesian interval construction.

\end{document}